\documentclass[final, times]{elsarticle}
\usepackage[letterpaper,top=2cm,bottom=2cm,left=3cm,right=3cm,marginparwidth=1.75cm]{geometry}
\usepackage{amsmath}
\usepackage{graphicx}
\usepackage{amssymb}
\usepackage{lipsum}
\usepackage{color}
\usepackage{xspace}
\usepackage{caption}
\usepackage{appendix}
\usepackage{float}
\usepackage{hyperref}
\usepackage{diagbox}
\usepackage{soul}
\usepackage{slashed}
\usepackage{amsfonts}
\usepackage{multirow}
\usepackage{mathrsfs}
\usepackage{bm}
\usepackage{bbm}
\usepackage{comment}
\usepackage{cuted}
\usepackage{setspace}
\biboptions{sort&compress}

\newcommand{\nn}{\nonumber}
\newcommand{\als}{\alpha_s}
\newcommand{\eps}{\epsilon}
\newcommand{\M}{\ensuremath \text{M}}
\begin{document}
\begin{frontmatter}

\title{Analytic Result of Higgs Boson Decay to Gluons with Full Quark Mass Dependence}

\author[a,b]{Jian Wang}
\author[c,d]{Yefan Wang}
\address[a]{School of Physics, Shandong University, Jinan, Shandong 250100, China}
\address[b]{Center for High Energy Physics, Peking University, Beijing 100871, China}
\address[c]{Department of Physics and Institute of Theoretical Physics, Nanjing Normal University, Nanjing,
Jiangsu 210023, China}
\address[d]{Nanjing Key Laboratory of Particle Physics and Astrophysics, Nanjing Normal University, Nanjing,
Jiangsu 210023, China}

\begin{abstract}
We report the first analytic calculation of the Higgs boson decay width to gluons up to next-to-leading order in quantum chromodynamics
including full dependence on the bottom and charm quark masses.
The interference between top- and bottom/charm-quark-induced amplitudes provides unexpectedly large contributions because the large logarithms of the bottom (charm) quark masses over the Higgs boson mass compensate for the suppression of the Yukawa coupling. 
These logarithms exhibit a unique structure distinct from Sudakov resummation and more complex than the Higgs-gluon-gluon form factor. 
Our results demonstrate that the previous calculation with massless bottom and charm quarks overestimates the decay width by nearly 20\%.
The prediction of the partial decay width with massive quarks provides essential theory input needed in the precise determination of Higgs couplings at future colliders.

\end{abstract}
\end{frontmatter}

\section{Introduction}
The discovery of the Higgs boson at the Large Hadron Collider (LHC) in 2012 \cite{ATLAS:2012yve,CMS:2012qbp} marked a triumph for the Standard Model (SM) and launched an era of precision Higgs physics.
The ATLAS and CMS collaborations at the LHC have provided constraints on Higgs couplings with uncertainties ranging from a few percent for couplings with massive gauge bosons to tens of percent for the couplings with massive fermions \cite{CMS:2022dwd,ATLAS:2022vkf,ATLAS:2024fkg}. 
Future colliders, such as the High-Luminosity LHC and proposed electron-positron colliders, will push these measurements to the (sub-)percent level \cite{ATLAS:2018jlh,Asner:2013psa}, demanding equally precise theoretical predictions to uncover potential beyond-SM physics.

Among the various decay channels of the Higgs boson, the decay to gluons, 
$H\to gg$, plays a particularly important role. 
It offers a direct measurement of the $Hgg$ coupling, complementary to the Higgs production cross section at the LHC.
In addition,  the digluon decay, combined with the $s$-channel Higgs production at an electron collider, is a very promising signal channel to measure the electron Yukawa coupling as it has the third most abundant branching fraction, and has no irreducible physical background, i.e., the process $e^+e^-\to Z/\gamma^*\to gg$ cannot happen \cite{dEnterria:2021xij,Abidi:2025dfw}.
Due to the clean environment, the signal strength of $H\to gg$ at future lepton colliders can be measured with $0.8\%$  precision \cite{Asner:2013psa,Tian:2016qlk,Freitas:2019bre,Zhu:2022lzv,CEPCPhysicsStudyGroup:2022uwl,selvaggi_2025_n2emg-43f06}.
These measurements would be useful only if combined with comparably accurate theory predictions.
This process occurs at the leading order (LO) through a loop of massive quarks, primarily the top quark due to its large Yukawa coupling. 
Theoretical predictions for this decay mode have been extensively studied.
The next-to-leading order (NLO) quantum chromodynamics (QCD) and electroweak corrections were calculated long ago \cite{Spira:1995rr,Actis:2008ug}.
When the top quark is taken to be infinitely heavy, the partial decay width has been calculated up to $\mathcal{O}(\alpha_s^6)$ \cite{Djouadi:1991tka,Chetyrkin:1997iv,Baikov:2006ch,Moch:2007tx,Herzog:2017dtz}.  
Meanwhile, the power-suppressed corrections in the inverse top-quark mass $m_t$ were investigated up to $\mathcal{O}(\alpha_s^4)$ \cite{Larin:1995sq,Schreck:2007um}. 
Consequently, the theoretical uncertainties from the scale variation and missing higher-order or higher-power corrections are reduced to below $1\%$.

Although the bottom-quark contribution is suppressed by $m_b^2/m_H^2\sim 0.1\%$ because of the Yukawa coupling and helicity conservation,
the corresponding decay amplitude receives strong logarithmic enhancements of the form $\log^2(m_b^2/m_H^2) \sim 40$,
making its interference with the top-quark-mediated amplitude quite sizable.
Indeed, as shown in Table \ref{tab:num} below, the interference can give rise to a correction of about $20\%$, 
which is of the same magnitude as the NNLO QCD correction to the top-quark-mediated decay process.
The large interference contribution highlights the necessity to include the physical quark mass effects in the prediction for the partial decay width of $H\to gg$.
These effects have been calculated analytically at LO and numerically at NLO in QCD \cite{Chetyrkin:1995pd,Spira:1995rr,Djouadi:1997yw}.
In this work, we are going to provide an analytic result of the NLO QCD correction involving contributions from finite bottom- and charm-quark masses,
which can generate accurate and fast predictions.

On the theory side, the logarithmic structure is of fundamental interest in its own right.
The logarithmic enhancements grow at higher orders in perturbation theory while the $m_b^2$ power suppression remains the same, i.e.,
the higher-order corrections in the bottom-quark-mediated amplitude turn out to be in the form of $\alpha_s^n \log^{2n}(m_b^2/m_H^2) $,
instead of the conventional $\alpha_s^n$.
Therefore, the traditional perturbative expansion would be invalidated if one considers the small quark mass limit.
 To provide precise and reliable theoretical predictions, it is crucial to perform resummation of such logarithms to all orders,
which is a nontrivial task because they are not the conventional Sudakov logarithms.
In fact, this kind of logarithm is induced by soft quarks, in contrast to soft gluons in Sudakov logarithms,
and has acquired considerable attention recently \cite{Liu:2017vkm,Moult:2019uhz,Liu:2019oav,Wang:2019mym,Beneke:2020ibj,Liu:2020tzd,vanBeekveld:2021mxn,Beneke:2022obx,Bell:2022ott}.
The resummation has been accomplished only for a limited number of processes and proves to be more complicated than Sudakov resummation because the relevant factorization formula, 
which is employed to disentangle the two scales in the logarithm, suffers from the problem of end-point divergences.
The elaborate resummation of large logarithms in the Higgs-gluon-gluon form factor has been realized in \cite{Liu:2018czl,Anastasiou:2020vkr,Liu:2022ajh}.
Nonetheless, our analytic result of the $H\to gg$ partial width, 
 including the real correction and phase space integration, 
reveals more intricate logarithmic structures.
It would help to develop a novel resummation formalism for more physical observables than form factors.

\section{Calculation framework and analytic results}
The $H\to gg $ decay process can be induced by two effective operators,
\begin{align}
\mathcal{O}_1 = 
\frac{H}{v}G_{\mu\nu}^a G^{a,\mu\nu}\,,\quad
\mathcal{O}_2= \frac{m_b}{v}H \bar{b} b+ \frac{m_c}{v} H \bar{c} c\, ,
\end{align}
with $v$ being the vacuum expectation value of the Higgs field.
The top-quark effects are incorporated in the corresponding Wilson coefficients, denoted by $C_1$ and $C_2$, respectively  \cite{Chetyrkin:1996ke,Chetyrkin:1997un,Liu:2015fxa,Davies:2017xsp}.

According to the combination structure of effective operators in the squared amplitudes, the decay width of $H\rightarrow gg$ can be decomposed into three parts, i.e.,
\begin{align}
\Gamma_{H\rightarrow gg}=  \Gamma^{C_1C_1}_{H\rightarrow gg} + \Gamma^{C_1C_2}_{H\rightarrow gg} + \Gamma^{C_2C_2}_{H\rightarrow gg}\,.
\end{align}
Each of them can be expanded in a series of the strong coupling $\als$, 
\begin{align}
\Gamma^{C_i C_j}_{H\rightarrow gg} &= C_i C_j \sum_n \left(\frac{\als}{\pi}\right)^n \Delta^{C_iC_j}_{n,gg}\,.
\label{Delta}
\end{align}
Note that $C_1$ starts from $\mathcal{O}(\alpha_s)$ 
and that the $\mathcal{O}_2$ operator contributes through loop diagrams.

\begin{figure}[ht]
	\centering
		\includegraphics[width=0.3\linewidth]{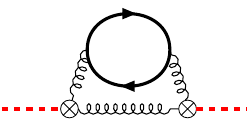}
		\includegraphics[width=0.3\linewidth]{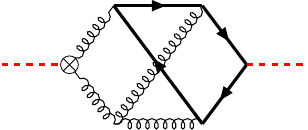}
		\includegraphics[width=0.3\linewidth]{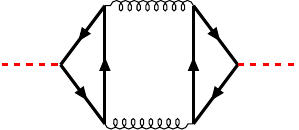}
\caption{Sample Feynman diagrams contributing to  $\Delta^{C_1C_1}_{i,gg}$ (left),  $\Delta^{C_1C_2}_{i,gg}$ (middle) and $\Delta^{C_2C_2}_{i,gg}$ (right).
 }
    \label{Diagrams}
\end{figure}

We compute the decay width of $H\rightarrow gg$ by the optical theorem, i.e.,
$
\Gamma_{H\rightarrow gg}  = \text{Im}_{gg}\left(\Sigma\right)/m_H  ,
$
where $\Sigma$ represents the amplitude of the process $H \rightarrow gg \rightarrow H$. 
The typical Feynman diagrams are shown in Fig. \ref{Diagrams}.
The notation $\text{Im}_{gg}$ denotes that only the imaginary part from a cut on at least two gluons or one gluon and one light-quark pair is retained.
The amplitudes with effective vertices were generated with {\sc FeynArts} \cite{Hahn:2000kx} 
and simplified with {\sc FeynCalc} \cite{Shtabovenko:2020gxv}.
The multiloop integrals were reduced to a set of master integrals using the {\sc Kira} package \cite{Klappert:2020nbg}.
We then applied the differential equation method \cite{Kotikov:1991pm,Henn:2013pwa} to calculate the master integrals.
The canonical master integrals in the calculation of the most complicated part, $\Delta_{2,gg}^{C_1 C_2}$, is shown explicitly in \ref{appendix}.
It is important to take into account the cut condition in every step.
In our calculation, the quark masses must be renormalized.
%We have performed renormalization in both the $\overline{\rm MS}$ and on-shell schemes.
In addition, it is only after including the renormalization mixing between the two effective operators that the ultraviolet divergences cancel out.
Interested readers are referred to our previous articles \cite{Wang:2023xud,Wang:2024ilc} for more details.

The result of the $C_1 C_1$ combination at NLO is given by
\begin{align}
\Delta^{C_1C_1}_{1,gg}&=  
\Delta^{C_1C_1}_{0,gg} \bigg\{\Big[-\frac{1}{3}l_z-\frac{1}{3}l_{z_c}-\frac{2}{3}l_{\mu} \Big]
+C_A\Big[
\frac{11}{6}l_{\mu}+\frac{73}{12}
\Big]+n_l\Big[-\frac{1}{3}l_{\mu}-\frac{7}{6}\Big]
\bigg\}
\end{align}
with $\Delta^{C_1C_1}_{0,gg}= C_AC_F m_H^3/ 2 \pi  v^2,l_z=\log(z),l_{z_c}=\log(z_c)$ and $l_{\mu}=\log\left(\mu ^2/m_H^2\right)$.
Here we define the dimensionless variable $z \equiv m_H^2/m_b^2$ with $m_b$ being the bottom-quark mass in the on-shell scheme.
A similar definition is applied to $z_c=m_H^2/m_c^2$.
The first term comes from the gluon self-energy with a massive quark loop.
We have used $n_l$ in the last term to denote the number of massless quark flavors.

The result of $\Delta^{C_1C_2}_{1,gg}$ is obtained by calculating the imaginary part of two-loop diagrams with one $\mathcal{O}_1$ and one $\mathcal{O}_2$ operators,
\begin{align}
\Delta^{C_1C_2}_{1,gg} &=\frac{m_Hm_b\overline{m_b}}{\pi v^2}C_AC_F\bigg\{\frac{z-4}{8 z}\Big[\log ^2\left(u\right)-\pi ^2\Big]-\frac{1}{2}\bigg\}
+(m_b \leftrightarrow m_c),
\end{align}
where the factor $\overline{m_b}$  comes from the bottom Yukawa coupling  renormalized in the $\overline{\rm MS}$ scheme 
while $m_b$ arises from the bottom quark propagator due to helicity conservation.
The argument $u=(z-2-\sqrt{z(z-4)})/2$ also depends on $m_b$. 
It is straightforward to express the result in terms of pure $\overline{\rm MS}$ or on-shell quark masses by making use of the relation between them; see, e.g., Refs. \cite{Melnikov:2000qh,Marquard:2015qpa}.

The evaluation of corresponding three-loop cut diagrams gives the following result.
\begin{align}
&\quad \Delta^{C_1C_2}_{2,gg} \,=\, \frac{m_Hm_b\overline{m_b}}{3\pi v^2}\frac{1}{(u+1)^2}\times
\nonumber\\   
&\bigg\{\bigg[27\operatorname{Li}_4(u^2)+72 \operatorname{Li}_4(u)-32\left(\operatorname{Li}_3(u^2)-\zeta (3)\right)\log(u)+\left(28\operatorname{Li}_2(-u)+16\operatorname{Li}_2(u)\right) \left(\log^2(u)-\pi ^2\right)
\bigg]
\nonumber\\&
\times\frac{(u-1)(u^2+1)}{2(u+1)}
+\frac{\left(13 u^4+54 u^3-72 u^2+54 u+5\right) \log^4(u)}{48 (u+1)^2}
\nonumber\\&
-\frac{\left(23 u^4+162 u^3-216 u^2+162 u+31\right)\pi ^2 \log^2(u)}{24 (u+1)^2}
-\frac{\left(94 u^4-540 u^3+720 u^2-540 u-274\right)\pi ^4 }{480 (u+1)^2}
\nonumber\\&
-4 (u-1)^2 [\operatorname{Li}_3(-u)+8\operatorname{Li}_3(u)-4 \operatorname{Li}_2(u)\log(u)]
+\left(51 u^2-110 u+51\right) [\operatorname{Li}_2(-u)\log(u)-\operatorname{Li}_3(-u)]
\nonumber\\&
+\frac{1}{2} \left(53 u^2-114 u+53\right) \left(\log^2(u)-\pi ^2\right)\log(u+1)
+\left(31 u^2-70 u+31\right) \zeta (3)
\nonumber\\&
-\frac{\left(141 u^3-205 u^2-7 u+39\right)\log^3(u)}{24 (u+1)}
+\frac{\left(335 u^3-527 u^2+67 u+29\right) \pi ^2\log(u)}{24 (u+1)}
\nonumber\\&
-\frac{\left(341 u^4+384 u^3-1588 u^2+384 u+341\right) \left(\pi ^2-\log^2(u)\right)}{16 (u+1)^2}
-12 (u+1)^2 \log(u+1)
\nonumber\\&
+\frac{3 \left(183 u^3-311 u^2+503 u-119\right)\log(u)}{16 (u+1)}-\frac{5285\left(u^2+1\right)}{32}-\frac{3611 u}{16}\bigg\}
\nonumber\\&
-\frac{1}{6}\left[\log(z)+\log\left(z_c \right)\right]\Delta^{C_1C_2}_{1,gg}+
(m_b\leftrightarrow m_c).
\end{align}
For brevity, the color factors $C_F=4/3$, $C_A=3$, $T_R=1/2, n_l=3$ and the renormalization scale $\mu =m_H$  have been substituted.
The full form can be found in the supplementary file.
It is remarkable that the three-loop cut amplitudes with multiple massive propagators can be expressed in terms of mere (poly)logarithmic functions.

The third kind of contribution, $\Delta^{C_2C_2}_{i,gg}$, comes from flavor singlet diagrams with two $\mathcal{O}_2$ operators.
The Yukawa couplings  and helicity conservation lead to an overall  factor $m_b^4 $, and thus the results are highly suppressed  compared to the first two kinds of contributions.
We present only the LO nonvanishing contribution below, 
\begin{align}
\Delta^{C_2C_2}_{2,gg} & = \frac{C_AC_F}{128 \pi v^2m_H}| F_b+F_c | ^2,\nonumber\\
F_b & = m_b\overline{m_b}\left[
\left(1-\frac{4}{z}\right) (\log(u)+i\pi)^2 - 4 \right],\, \nonumber \\
F_c & = F_b|_{m_b\rightarrow m_c}\,.
\end{align}

In the small quark mass limit, the results in the $C_1 C_2$ and $C_2 C_2$ combinations can be expanded as follows.
\begin{align}
&\Delta^{C_1C_2}_{1,gg}|_{z \rightarrow \infty} = \frac{m_Hm_b\overline{m_b}}{\pi v^2}C_AC_F
\bigg[\frac{1}{8}(l_z^2-\pi^2-4) -\frac{1}{2z}(l_z^2+l_z-\pi^2)+\mathcal{O}(z^{-2})\bigg]\,,
\label{eq:c1c2}\\
&\Delta^{C_2C_2}_{2,gg}|_{z \rightarrow \infty}  = \frac{m_b^2 \overline{m_b}^2}{128\pi v^2 m_H}C_AC_F\bigg[\big(l_z^2-4 l_z
 +\pi ^2+4\big) \left(l_z^2+4 l_z+\pi ^2+4\right)+\mathcal{O}(z^{-1})\bigg]\,.
\label{eq:c2c2}
\end{align}
Here and below, for simplicity, we omit the charm-quark contribution in the expansion, which has a similar structure.
In the method of regions \cite{Beneke:1997zp}, the above logarithms arise from the regions where the internal bottom quark has a soft momentum or a momentum collinear to the external gluons.
The end-point divergences appear in each of the two regions, but cancel with each other \cite{Liu:2019oav,Hou:2025ovb}.

The NLO correction in the $C_1 C_2$ combination is expanded as
\begin{align}
&\Delta^{C_1C_2}_{2,gg}|_{z \rightarrow \infty} = \frac{m_Hm_b\overline{m_b}}{\pi v^2}C_AC_F\times\nonumber\\
&\bigg\{
C_A\Big[\frac{1}{192}l_z^4+\frac{13}{288}l_z^3-\Big(\frac{\pi ^2}{32}-\frac{349}{576}\Big)l_z^2
-\Big(\frac{13 \pi ^2}{96}-\frac{115}{192}\Big)l_z+\frac{\pi ^4}{192}+\frac{11 \zeta (3)}{12}
-\frac{349 \pi ^2}{576}-\frac{5321}{1152}
\Big]
\nonumber\\&
+C_F\Big[
-\frac{1}{192}l_z^4+\frac{1}{32}l_z^3-\Big(\frac{\pi ^2}{96} -\frac{1}{8} \Big) l_z^2
+\Big(\zeta (3)+\frac{13 \pi ^2}{96}+\frac{3}{8}\Big)l_z+\frac{23 \pi ^4}{960}
+\frac{\zeta (3)}{4}-\frac{\pi ^2}{8}-\frac{7}{4}
\Big]
\nonumber\\&
+n_l\Big[
-\frac{1}{72}l_z^3-\frac{5}{72}l_z^2+\Big(\frac{\pi ^2}{24}-\frac{7}{48}\Big)l_z
-\frac{\zeta (3)}{6}+\frac{5 \pi ^2}{72}+\frac{233}{288}
\Big]
+\Big[
-\frac{1}{24} l_z^2+\Big(\frac{\pi ^2}{24}+\frac{1}{6}\Big)
\Big]l_z+\mathcal{O}(z^{-1})
\bigg\}\,,
\label{eq:c1c2nlo}
\end{align}
where we retain all color structures explicitly.
The ratio of the leading logarithms in $\Delta^{C_1C_2}_{2,gg}$ and $\Delta^{C_1C_2}_{1,gg}$ reads 
$  (C_A-C_F)\log^2(z)/24 $.
This is rather different from conventional Sudakov double logarithms, which involve only $C_A$ or $C_F$.
The appearance of the color structure $C_A-C_F$ is a unique feature of the soft quark effect,
reflecting nonconservation of the color charge along a collinear direction,
and has been discovered in different contexts, e.g., the off-diagonal splitting functions \cite{Vogt:2010cv,Almasy:2010wn,Beneke:2020ibj}
and the thrust distribution in $e^+ e^-$ annihilation \cite{Moult:2019uhz,Beneke:2022obx}.
It is interesting to compare our result to the $Hgg$ form factor in \cite{Liu:2018czl,Liu:2022ajh}.
The form factor is unphysical and contains infrared divergences, which can be written in a factorized form due to its universality.
The leading logarithms in our result agree with the finite part of the $Hgg$ form factor if the infrared divergences are stripped off at a hard scale.
However, the next-to-leading logarithms are very different, 
since the real corrections, i.e., $H\to ggg (gq\bar{q}) $ decay processes, could also contribute to large logarithms.
At higher power in $m_b^2$, even the leading logarithms does not coincide with the result of the form factor in \cite{Liu:2021chn}.
Therefore, novel factorization formulae need to be developed to resum the large logarithms in the $H\to gg$ partial decay width.
Our analytic result presented above provides a useful guideline.

Summarizing the above results, we obtain the  partial decay width for $H\to gg$ up to $\alpha_s^3$.
Combining with the result of $H\to b\bar{b}$ in our previous papers \cite{Wang:2023xud,Wang:2024ilc} yields the hadronic decay width, 
which is consistent with the results at leading power of $m_b^2$ in
\cite{Chetyrkin:1997vj,Davies:2017xsp}.
This serves as a valuable validation of our calculation.

In the above calculations, we have already taken the limit of $m_t\to \infty$ and retained the leading power of $x \equiv m_H^2/m_t^2$.
We compute the next-to-leading power correction, obtaining
% This mt is on-shell mt, we need to transform the MSBar mt to on-shell mt from 0708.0916 eq.6%
\begin{align}
&\Gamma_{H\rightarrow gg}^x = 
\frac{m_H^3 x}{v^2\pi }\bigg\{\left(\frac{\als}{\pi}\right)^2\bigg[\frac{7}{4320} + \frac{1}{{z}} \bigg(-\frac{7}{2880}l_z^2+\frac{7 \pi ^2}{2880}+\frac{7}{720}\bigg)
+\mathcal{O}({z}^{-2}) \bigg] \label{eq:xcor}\\
&+\left(\frac{\als}{\pi}\right)^3\bigg[
\frac{-7}{12960}\left(l_z+l_{\mu}\right)+\frac{1423}{38880}+\frac{7}{960}l_\mu+\mathcal{O}({z}^{-1})\bigg]\bigg\},\nonumber
\end{align}
%where we have chosen $\mu =m_H$. 
where the $\mathcal{O}(\als^3)$ correction is new, while the $\mathcal{O}(\als^2)$ correction can be found in \cite{Larin:1995sq,Chetyrkin:1995pd}.

\section{Numerical results}
We calculate the decay width numerically by taking the input parameters
\begin{align}
    \overline{m_b}\,(\overline{m_b}) &= 4.18 ~{\rm GeV}, \quad  \overline{m_c}\,(\overline{m_c}) = 1.27 ~{\rm GeV},  \nonumber\\
     m_H & = 125.09 ~{\rm GeV},\quad  
     m_t  = 172.69 ~{\rm GeV}, \\
     \als(m_Z) & = 0.1181,  ~~G_F  = 1.166378 \times 10^{-5}~{\rm GeV}^{-2}\,. \nn
    \label{eq:input}    
\end{align}
The conversion between the $\overline{\rm MS}$ and on-shell quark masses is performed using the package {\tt RunDec} \cite{Chetyrkin:2000yt,Herren:2017osy} with four-loop relations \cite{Marquard:2015qpa}.
%and strong coupling at other scales are determined 
%e.g.,  $\overline{m_b}\,(m_H) = 2.78425$ GeV、$\overline{m_c} = 0.59630$ GeV\and $\als\,(m_H) = 0.112715$. 
The contributions from various coupling combinations to the partial decay width in the $\overline{\rm MS}$ and on-shell schemes are shown in Table \ref{tab:num}.

At $\mathcal{O}(\alpha_s^2)$, the $C_1 C_1$ combination is dominant,
while the $C_1 C_2$ combination reduces the decay with by $20\%$ ($7\%$) in the on-shell ($\overline{\rm MS}$) scheme. 
The correction from the $C_2 C_2$ combination in the on-shell ($\overline{\rm MS}$) scheme only amounts to $3\%$  ($0.3\%$) of that from $C_1 C_1$.
This hierarchy in the magnitude does not follow the power counting in the ratio of quark masses to the Higgs boson mass. 
An enhancement from the logarithmic terms should be taken into account to explain the relative size; see Eqs. (\ref{eq:c1c2},\ref{eq:c2c2}).
Since the contribution from the $C_2 C_2$ combination is so small, it is expected that its higher-order corrections are negligible in phenomenological studies.

The NLO QCD correction in the $C_1 C_1$ combination increases the LO result by around $56\%$ and $51\%$ in the on-shell  and $\overline{\rm MS}$ schemes, respectively.
The notable correction is not unexpected given that the QCD corrections in hadronic Higgs production at the LHC, which is related to $H\to gg$ decay by crossing symmetry, are also significant \cite{Anastasiou:2015vya}.
This behavior indicates that the NNLO QCD corrections are probably still important.
The result for the $H\to gg$ decay with massless bottom quarks was computed a long time ago, and a correction of $20\%$ is found at NNLO \cite{Chetyrkin:1997iv,Baikov:2006ch}.
However, the result with massive bottom quarks is still missing.
The NLO QCD correction in the $C_1 C_2$ combination is about $42\%$ ($83\%$) of the corresponding LO result in the on-shell ($\overline{\rm MS}$) scheme.
The dominant corrections come from the logarithms of the quark masses in Eq.~(\ref{eq:c1c2nlo}).

\begin{figure}[ht]
	\centering
\includegraphics[width=0.6\linewidth]{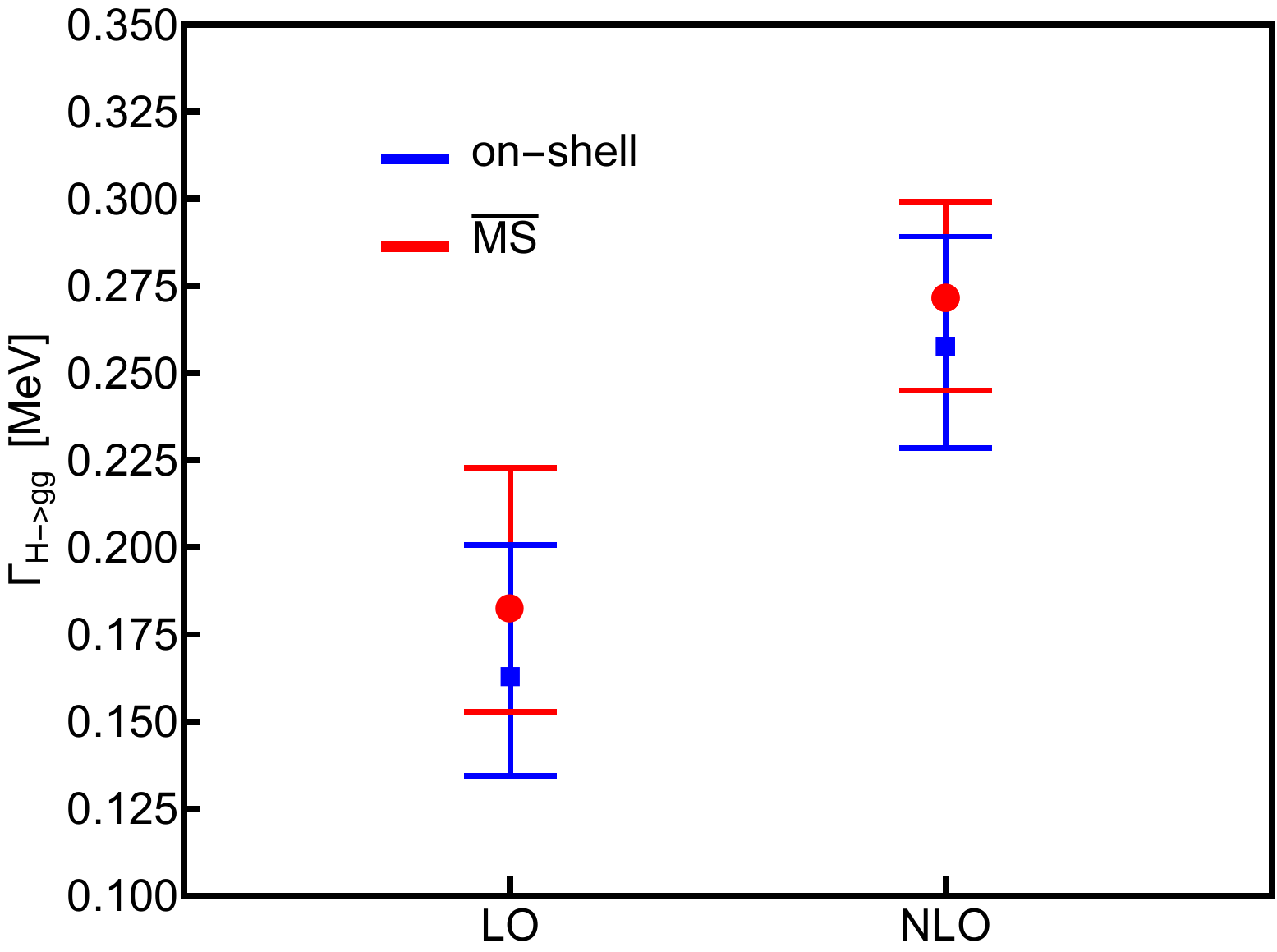}
 \caption{The full $\Gamma_{H\rightarrow gg}$  at different orders of $\als$ in the on-shell and $\overline{\rm MS}$ schemes for the quark masses.}
 \label{fig:scheme} 
\end{figure}

We show the full $H\to gg$ decay widths at LO and NLO in Fig.  \ref{fig:scheme},
which include the power corrections of $m_H^2/m_t^2$ in Eq. (\ref{eq:xcor}) (around $7\%$ numerically).
The  scale  uncertainty changes from 24\% at LO to 12\% at NLO.
The difference between the results in the on-shell and ${\rm \overline{MS}}$ schemes is reduced by a factor of two after including the NLO QCD correction.
Explicitly, we obtain
\begin{subequations}
\begin{align}
    \Gamma_{H\to gg}^{\rm NLO}~({\textrm{on-shell}}) & = 0.257 ^{+0.032}_{-0.029} ~\textrm{MeV}\,,\\
    \Gamma_{H\to gg}^{\rm NLO}~({\rm \overline{MS}}) &= 0.272 ^{+0.027}_{-0.027} ~\textrm{MeV}\,, 
    \label{eq:Gammamsbar}
\end{align}
\end{subequations}
where the errors denote the scale uncertainties.
We emphasize that the above prediction in the on-shell (${\rm \overline{MS}}$) scheme is lower than the result with massless bottom and charm quarks \cite{Schreck:2007um} by  $20\%$ ($15\%$)\footnote{
In Ref. \cite{Schreck:2007um}, only $\Gamma_{H\to gg}^{C_1C_1}$ is considered, in which the value of $n_l$ is taken to be different.
Moreover, it includes in $\Gamma_{H\to gg}^{C_1C_1}$ the contribution from the decay process $H\to gb\bar{b}$ to obtain a finite result, since the bottom quark is assumed to be massless.
By contrast, this decay process is classified as the contribution to $H\to b\bar{b}$ in our work because the massive bottom quark, after forming a B meson, can be detected using the reconstructed secondary vertices.
Therefore, taking the massless limit of our prediction on $\Gamma_{H\to gg}$ does not reproduce the result in Ref. \cite{Schreck:2007um}. 
}, indicating that 
it is essential to use the results with physical quark masses to confront the experimental data in future.

\begin{table*}[]
	\centering
		\begin{tabular}{cc|ccc|ccc}
			\hline
			& & \multicolumn{3}{c|}{On-shell scheme} & \multicolumn{3}{c}{$\overline{\rm MS}$ scheme} \\
			\hline
			& [MeV] & $\mu = m_H/2$ & $\mu = m_H$ & $\mu = 2m_H$ & $\mu = m_H/2$ & $\mu = m_H$ & $\mu = 2m_H$ \\
			\hline
			\multirow{3}{*}{$\mathcal{O}(\alpha_s^2)$} 
			& $\Gamma^{C_1C_1}_{H\to gg}$ & 0.2269 & 0.1837 & 0.1520 & 0.2269 & 0.1837 & 0.1520 \\
			& $\Gamma^{C_1C_2}_{H\to gg}$ & -0.0449 & -0.0364 & -0.0301 & -0.0177 & -0.0133 & -0.0102 \\
			& $\Gamma^{C_2C_2}_{H\to gg}$ & 0.0063 & 0.0051 & 0.0042 & 0.0007 & 0.0005 & 0.0004 \\
			\hline
			\multirow{2}{*}{$\mathcal{O}(\alpha_s^3)$} 
			& $\Gamma^{C_1C_1}_{H\to gg}$ & 0.0928 & 0.1027 & 0.1036 & 0.0807 & 0.0933 & 0.0961 \\
			& $\Gamma^{C_1C_2}_{H\to gg}$ & -0.0113 & -0.0152 & -0.0166 & -0.0111 & -0.0110 & -0.0102 \\
			\hline
		\end{tabular}
	\caption{The contributions $\Gamma^{C_2C_2}_{H\rightarrow gg}$, $\Gamma^{C_1C_2}_{H\rightarrow gg}$ and $\Gamma^{C_1C_1}_{H\rightarrow gg}$  at different orders of $\als$ in the on-shell and $\overline{\rm MS}$ schemes for the quark masses.} 
	\label{tab:num}
\end{table*}

\section{Conclusion}
We have calculated the $H\to gg$ decay width with full dependence on the bottom- and charm-quark masses analytically.
The asymptotic expansion in the small quark mass limit exhibits intriguing structures that are different from conventional Sudakov logarithms and those in the Higgs-gluon-gluon form factor.
Our numerical results show that the massive bottom- and charm-quark-induced contributions are quite sizable due to large logarithms of the quark mass to the Higgs boson mass. 
The NLO QCD corrections change the decay width significantly, 
with the renormalization scale and scheme dependencies reduced by half.
These results with physical quark masses, differing from the previous massless approximation by about 20\%,
are indispensable in precision tests of the SM in the Higgs sector at future colliders.
Our research can be extended in several directions. 
The NNLO QCD corrections in the presence of quark masses would be needed to further reduce the theoretical uncertainties.
It is also interesting to understand the intriguing logarithmic structures revealed in the analytic result of the decay width, which may stimulate the development of a novel resummation formalism.

\section*{Acknowledgements}
We are grateful to Xunwu Zuo for pointing out the importance of the Higgs digluon decay in the determination of the electron Yukawa coupling.
This work was partially supported by the National Natural Science Foundation of China under grants  No. 12321005, No. 12375076, and  No.12405117. J.W. was also supported by the Taishan Scholar Foundation of Shandong province (tsqn201909011)

\appendix
\section{Master integrals in the calculation of $\Delta^{C_1C_2}_{2,gg}$}
\label{appendix}
\begin{figure}[h]
    \centering
    \includegraphics[width=0.4\linewidth]{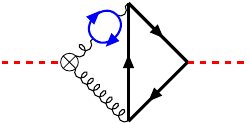}
    \caption{The three-loop diagram containing both bottom and charm quarks. The thick black, red and blue lines stand for the bottom quark, Higgs boson and charm quark, respectively.}
    \label{fig:bc}
\end{figure}

Firstly, we consider the three-loop diagrams containing both bottom and charm quarks. The typical Feynman diagram is shown in Fig. \ref{fig:bc}.
This contribution can be easily obtained by multiplying $\Delta^{C_1C_2}_{1,gg}$ by the renormalization constant of the gluon wave function. Thus, we need to calculate the three-loop integrals with only one massive quark.

\begin{figure}[ht]
    \centering
    \includegraphics[width=0.3\linewidth]{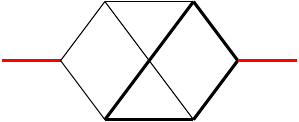}
    \caption{The topology of the NP1 integral families. The thick black and red lines
 stand for the massive bottom (charm) quark and the Higgs boson, respectively. The other lines denote the gluons.}
    \label{fig:NP1}
\end{figure}

\begin{figure}[ht]
	\centering
	\begin{minipage}{0.2\linewidth}
		\centering
		\includegraphics[width=1\linewidth]{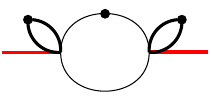}
		\caption*{$\M^{\text{NP1}}_{1}$}
	\end{minipage}
	\begin{minipage}{0.2\linewidth}
		\centering
		\includegraphics[width=1\linewidth]{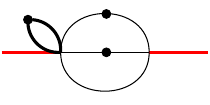}
		\caption*{$\M^{\text{NP1}}_{2}$}
	\end{minipage}
	\begin{minipage}{0.2\linewidth}
		\centering
		\includegraphics[width=1\linewidth]{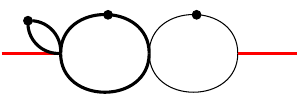}
		\caption*{$\M^{\text{NP1}}_{3}$}
	\end{minipage}
	\centering
	\begin{minipage}{0.2\linewidth}
		\centering
		\includegraphics[width=1\linewidth]{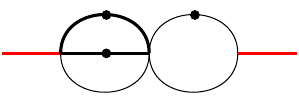}
		\caption*{$\M^{\text{NP1}}_{4}$}
	\end{minipage}
	\begin{minipage}{0.2\linewidth}
		\centering
		\includegraphics[width=1\linewidth]{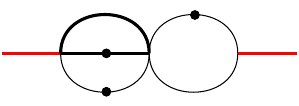}
		\caption*{$\M^{\text{NP1}}_{5}$}
	\end{minipage}
	\begin{minipage}{0.2\linewidth}
		\centering
		\includegraphics[width=1\linewidth]{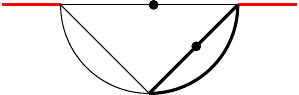}
		\caption*{$\M^{\text{NP1}}_{6}$}
	\end{minipage}
	\begin{minipage}{0.2\linewidth}
		\centering
		\includegraphics[width=1\linewidth]{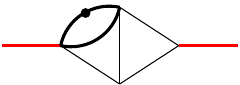}
		\caption*{$\M^{\text{NP1}}_{7}$}
	\end{minipage}
	\begin{minipage}{0.2\linewidth}
		\centering
		\includegraphics[width=1\linewidth]{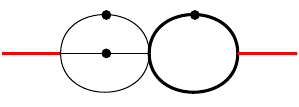}
		\caption*{$\M^{\text{NP1}}_{8}$}
	\end{minipage}
	\begin{minipage}{0.2\linewidth}
		\centering
		\includegraphics[width=1\linewidth]{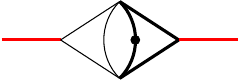}
		\caption*{$\M^{\text{NP1}}_{9}$}
	\end{minipage}
	\begin{minipage}{0.2\linewidth}
		\centering
		\includegraphics[width=1\linewidth]{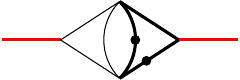}
		\caption*{$\M^{\text{NP1}}_{10}$}
	\end{minipage}
	\begin{minipage}{0.2\linewidth}
		\centering
		\includegraphics[width=1\linewidth]{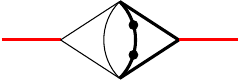}
		\caption*{$\M^{\text{NP1}}_{11}$}
	\end{minipage}
	\begin{minipage}{0.2\linewidth}
		\centering
		\includegraphics[width=1\linewidth]{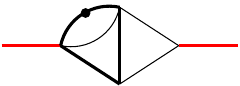}
		\caption*{$\M^{\text{NP1}}_{12}$}
	\end{minipage}
	\begin{minipage}{0.2\linewidth}
		\centering
		\includegraphics[width=1\linewidth]{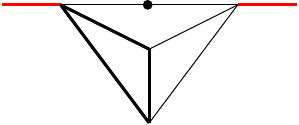}
		\caption*{$\M^{\text{NP1}}_{13}$}
	\end{minipage}
	\begin{minipage}{0.2\linewidth}
		\centering
		\includegraphics[width=1\linewidth]{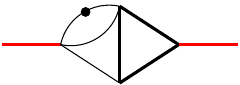}
		\caption*{$\M^{\text{NP1}}_{14}$}
	\end{minipage}
	\begin{minipage}{0.2\linewidth}
		\centering
		\includegraphics[width=1\linewidth]{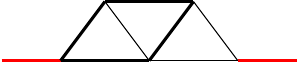}
		\caption*{$\M^{\text{NP1}}_{15}$}
	\end{minipage}
	\begin{minipage}{0.2\linewidth}
		\centering
		\includegraphics[width=1\linewidth]{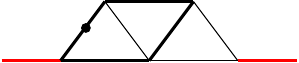}
		\caption*{$\M^{\text{NP1}}_{16}$}
	\end{minipage}
	\begin{minipage}{0.2\linewidth}
		\centering
		\includegraphics[width=1\linewidth]{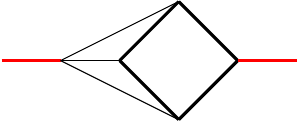}
		\caption*{$\M^{\text{NP1}}_{17}$}
	\end{minipage}
	\begin{minipage}{0.2\linewidth}
		\centering
		\includegraphics[width=1\linewidth]{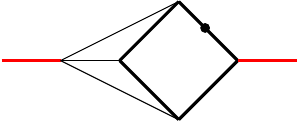}
		\caption*{$\M^{\text{NP1}}_{18}$}
	\end{minipage}
	\begin{minipage}{0.2\linewidth}
		\centering
		\includegraphics[width=1\linewidth]{M19.pdf}
		\caption*{$\M^{\text{NP1}}_{19}$}
	\end{minipage}
\caption{Master integrals in the NP1 topology. The thick black and red lines stand for the bottom (charm) quark and the Higgs boson, respectively. One black dot indicates one additional power of the corresponding propagator.}
\label{NP1_Topo}
\end{figure}

After considering the symmetries among integrals with the help of the package {\tt CalcLoop}\footnote{\url{https://gitlab.com/multiloop-pku/calcloop}}, most of the master integrals can be classified into one integral family, denoted by NP1, which is shown in Fig. \ref{fig:NP1}. The other master integrals can be factorized into one- and two-loop integrals which are easy to compute.  
The integrals in the NP1 family are represented by 
\begin{align}
	I^{\rm NP1}_{n_1,n_2,\ldots,n_{9}}=\textrm{Im}_{gg} \int \prod_{i=1}^3  \frac{(m_Q^2)^{\eps}d^D q_i}{i\pi^{D/2}\Gamma(1+\eps)}~\frac{D_9^{-n_9}}{D_1^{n_1}~D_2^{n_2}~D_3^{n_3}~D_4^{n_4}~D_5^{n_5}~D_6^{n_6}~D_7^{n_7}D_8^{n_8}}, 
\end{align}
with $D=4-2\eps$ and
all $n_i\geqslant 0,i=1,\cdots, 8$,  $n_9\leqslant 0$. 
The denominators $D_i$ read
\begin{align}
	D_1 &= q_1^2,&
	D_2 &= (q_1-q_2)^2-m_Q^2,&
	D_3 &= q_2^2-m_Q^2,\nonumber\\
	D_4 &= q_3^2-m_Q^2,&
	D_5 &= (q_2+q_3)^2,&
	D_6 &= (q_3-k)^2-m_Q^2,\nonumber\\
	D_7 &= (q_1+k)^2,&
	D_8 &=(q_1-q_2-q_3+k)^2,&
	D_9 &=(q_2-k)^2,
\end{align}
The external momentum $k$ satisfies $k^2 = m_H^2$. In the above definition of integrals, we have required only the imaginary contribution induced by a cut on at least two gluon lines. Meanwhile, the cuts on massive (bottom or charm) quarks need to be excluded. In this way there are 19 master integrals in this family after the integration by parts (IBP) reduction \cite{Tkachov:1981wb,Chetyrkin:1981qh}. The topology diagrams of the master integrals in the NP1 families are displayed in Fig. \ref{NP1_Topo}.

To construct the canonical basis, we first select dimensionless integrals
\begin{align}
\M^{\text{NP1}}_{1}& = \epsilon ^3 I^{\text{NP1}}_{2,2,0,2,0,0,1,0,0} m_Q^2,\quad&
\M^{\text{NP1}}_{2}& = \epsilon ^3 I^{\text{NP1}}_{2,2,0,0,2,0,0,1,0} m_Q^2,\quad\nonumber\\
\M^{\text{NP1}}_{3}& = \epsilon ^3 I^{\text{NP1}}_{2,2,0,2,0,1,1,0,0} m_Q^4,\quad&
\M^{\text{NP1}}_{4}& = \epsilon ^3 I^{\text{NP1}}_{2,0,2,0,1,2,1,0,0} m_Q^4,\quad\nonumber\\
\M^{\text{NP1}}_{5}& = \epsilon ^3 I^{\text{NP1}}_{2,0,2,0,2,1,1,0,0} m_Q^4,\quad&
\M^{\text{NP1}}_{6}& = (1-2 \epsilon ) \epsilon ^3 I^{\text{NP1}}_{1,2,0,1,1,0,0,2,0} m_Q^2,\quad\nonumber\\
\M^{\text{NP1}}_{7}& = (1-2 \epsilon ) \epsilon ^4 I^{\text{NP1}}_{1,2,0,1,1,0,1,1,0} m_Q^2,\quad&
\M^{\text{NP1}}_{8}& = \epsilon ^3 I^{\text{NP1}}_{2,0,0,2,2,1,0,1,0} m_Q^4,\quad\nonumber\\
\M^{\text{NP1}}_{9}& = (1-2 \epsilon ) \epsilon ^4 I^{\text{NP1}}_{1,2,0,1,1,1,1,0,0} m_Q^2,\quad&
\M^{\text{NP1}}_{10}& = (1-2 \epsilon ) \epsilon ^3 I^{\text{NP1}}_{1,2,0,2,1,1,1,0,0} m_Q^4,\quad\nonumber\\
\M^{\text{NP1}}_{11}& = (1-2 \epsilon ) \epsilon ^3 I^{\text{NP1}}_{1,3,0,1,1,1,1,0,0} m_Q^4,\quad&
\M^{\text{NP1}}_{12}& = (1-2 \epsilon ) \epsilon ^4 I^{\text{NP1}}_{1,1,1,0,1,2,1,0,0} m_Q^2,\quad\nonumber\\
\M^{\text{NP1}}_{13}& = (1-2 \epsilon ) \epsilon ^4 I^{\text{NP1}}_{1,1,1,1,1,0,0,2,0} m_Q^2,\quad&
\M^{\text{NP1}}_{14}& = (1-2 \epsilon ) \epsilon ^4 I^{\text{NP1}}_{2,1,0,1,1,1,0,1,0} m_Q^2,\quad\nonumber\\
\M^{\text{NP1}}_{15}& = (1-2 \epsilon ) \epsilon ^5 I^{\text{NP1}}_{1,1,1,1,1,1,1,0,0} m_Q^2,\quad&
\M^{\text{NP1}}_{16}& = (1-2 \epsilon ) \epsilon ^4 I^{\text{NP1}}_{1,1,1,2,1,1,1,0,0} m_Q^4,\quad\nonumber\\
\M^{\text{NP1}}_{17}& = (1-2 \epsilon ) \epsilon ^5 I^{\text{NP1}}_{1,1,1,1,1,1,0,1,0} m_Q^2,\quad&
\M^{\text{NP1}}_{18}& = (1-2 \epsilon ) \epsilon ^4 I^{\text{NP1}}_{1,1,1,2,1,1,0,1,0} m_Q^4,\quad\nonumber\\
\M^{\text{NP1}}_{19}& = (1-2 \epsilon ) \epsilon ^5 I^{\text{NP1}}_{1,1,1,1,1,1,1,1,0} m_Q^4.\quad&    
\end{align}
Then we obtain the following canonical basis of the NP1 family: 
\begin{align}
F^{\text{NP1}}_1 &= -z \M_1^{\text{NP1}},\quad
F^{\text{NP1}}_2 =-z \M_2^{\text{NP1}},\quad
F^{\text{NP1}}_3 =-r z \M_3^{\text{NP1}},\quad
F^{\text{NP1}}_4 =z^2 \M_4^{\text{NP1}},\quad
F^{\text{NP1}}_5 =\frac{1}{2} r z \M_4^{\text{NP1}}+r z \M_5^{\text{NP1}},\quad\nonumber\\
F^{\text{NP1}}_6 &=-\frac{1}{2} r z \M_4^{\text{NP1}}+r \M_6^{\text{NP1}}+r \M_7^{\text{NP1}},\quad
F^{\text{NP1}}_7 =-z \M_7^{\text{NP1}},\quad
F^{\text{NP1}}_8 =-r z \M_8^{\text{NP1}},\quad
F^{\text{NP1}}_9 =-z \M_9^{\text{NP1}},\quad\nonumber\\
F^{\text{NP1}}_{10} &=\frac{3 r \M_9^{\text{NP1}}}{2}+r \M_{10}^{\text{NP1}}+r \M_{11}^{\text{NP1}},\quad
F^{\text{NP1}}_{11} =  -z \M_{11}^{\text{NP1}},\quad
F^{\text{NP1}}_{12} =-z \M_{12}^{\text{NP1}},\quad
F^{\text{NP1}}_{13} =-z \M_{13}^{\text{NP1}},\quad\nonumber\\
F^{\text{NP1}}_{14} &=-z \M_{14}^{\text{NP1}},\quad
F^{\text{NP1}}_{15} =-z \M_{15}^{\text{NP1}},\quad
F^{\text{NP1}}_{16} = \frac{r z^2 \M_4^{\text{NP1}}}{4 (2-2 z)}+\frac{r z \M_9^{\text{NP1}}}{2 (z-2)}+\frac{r z \M_{11}^{\text{NP1}}}{z-2}+\frac{r z \M_{12}^{\text{NP1}}}{2 (2-z)}+\frac{r z \M_{15}^{\text{NP1}}}{2-z}+\frac{r z \M_{16}^{\text{NP1}}}{2-z},\quad\nonumber\\
F^{\text{NP1}}_{17} &= -z \M_{17}^{\text{NP1}},\quad
F^{\text{NP1}}_{18} = \frac{r \M_{13}^{\text{NP1}}}{2}-\frac{r \M_{14}^{\text{NP1}}}{2}+r \M_{17}^{\text{NP1}}+r \M_{18}^{\text{NP1}},\quad
F^{\text{NP1}}_{19} = z^2 \M_{19}^{\text{NP1}}\,,
\end{align}
with 
$
r = \sqrt{z(z-4)}.
$
The analytic results of the canonical basis mentioned above can be expressed in terms of multiple polylogarithms (MPL) \cite{Goncharov:1998kja}, which are then converted to polylogarithms. We provide the analytical results in the auxiliary files submitted along this paper.
\bibliographystyle{JHEP} 
\bibliography{reference.bib}

\end{document}